\documentclass[journal=nalefd]{achemso}

\usepackage{graphicx}    %Importing graphics with \includegraphics
\usepackage{amsmath}	% Better equation environment 
\usepackage{bm}				% Bold face for vectors
\usepackage{tabularx}	    % Control table width
\usepackage{footmisc}		% Better footnote capabilities

\newcommand{\beginsupplement}{%
        \setcounter{table}{0}
        \renewcommand{\thetable}{S\arabic{table}}%
        \setcounter{figure}{0}
        \renewcommand{\thefigure}{S\arabic{figure}}%
        \setcounter{equation}{0}
        \renewcommand{\theequation}{S\arabic{equation}}%
     }

\author{Jason W Christopher}
\email{jwc@bu.edu}
\affiliation[Department of Physics, Boston University]{Department of Physics, Boston University, 590 Commonwealth Ave, Boston, Massachusetts 02215, USA}

\author{Bennett B Goldberg}
\email{goldberg@bu.edu}
\affiliation[Department of Physics, Boston University]{Department of Physics, Boston University, 590 Commonwealth Ave, Boston, Massachusetts 02215, USA}
\alsoaffiliation[Nanotechnology Innovation Center, Boston University]{Nanotechnology Innovation Center, Boston University, 8 St Mary's St, Boston, Massachusetts 02215, USA}
\alsoaffiliation[Boston University Photonics Center]{Photonics Center, Boston University, 8 St Mary's St, Boston, Massachusetts 02215, USA}

\author{Anna K Swan}
\email{swan@bu.edu}
\affiliation[Department of Electrical and Computer Engineering, Boston University]{Department of Electrical and Computer Engineering, Boston University, 8 St Mary's St, Boston, Massachusetts 02215, USA }
\alsoaffiliation[Department of Physics, Boston University]{Department of Physics, Boston University, 590 Commonwealth Ave, Boston, Massachusetts 02215, USA}
\alsoaffiliation[Boston University Photonics Center]{Photonics Center, Boston University, 8 St Mary's St, Boston, Massachusetts 02215, USA}

\title[]{Long tailed trions in monolayer MoS$_2$: Temperature dependent asymmetry and red-shift of trion photoluminescence spectra}


\begin{document}
\clearpage %To keep abstract all on the same page

%%%%%%%%%%%%%%%%%%%%%%%%%%%%
%% Abstract (< 4000 Characters) 
%%%%%%%%%%%%%%%%%%%%%%%%%%%%
\begin{abstract}
Monolayer molybdenum disulfide (MoS$_2$) has emerged as an excellent 2D model system because of its two inequivalent, direct-gap valleys that lead to exotic bound and excited states. Here we focus on one such bound state, the negatively charged trion. Unlike excitons, trions can radiatively decay with non-zero momentum by kicking out an electron, resulting in an asymmetric trion photoluminescence (PL) peak with a long low-energy tail.  As a consequence, the peak position does not correspond to the zero momentum trion energy. By including the trion's long tail in our analysis we are able to accurately separate the exciton from the trion contributions to the PL spectra. According to theory, the asymmetric energy tail has  both a size-dependent and a temperature-dependent contribution. Analysis of the temperature-dependent data reveals the effective trion size, consistent with literature, and the temperature dependence of the band gap and spin-orbit splitting of the valence band.  Finally, we observe signatures of Pauli-blocking of the trion decay.
\end{abstract}

%%%%%%%%%%%%%%%%%%%%%%%%%%%%
%% Main part of the manuscript.
%%%%%%%%%%%%%%%%%%%%%%%%%%%%

The two-dimensionality of MoS$_2$ naturally reduces dielectric screening, resulting in strong interactions and exotic many-body bound excited states such as trions\cite{Mak2013a} and bi-excitons\cite{Sie2015a}.  The binding energies of these states in MoS$_2$ are nearly an order of magnitude larger than in GaAs quantum wells (QW)\cite{Esser2000}, which for trions in MoS$_2$ is large enough to make them stable even at room temperature\cite{Mak2013a}.  Like other Transition Metal Dichalcogenides (TMDC), MoS$_2$'s band structure contains two direct-gap inequivalent valleys with identical bands but with opposite spins due to time-reversal symmetry\cite{Xu2014}.  This symmetry makes it possible to optically address excitations at a specific valley, or coherently generate excitations between valleys\cite{Mak2012,Mak2014a,Zeng2012a}.  Effectively, the single particle states are endowed with a pseudo-spin degree of freedom called the valley index, which remarkably continues to be conserved in more complicated many-body states.  There is great interest in exciting and manipulating these states to further our understanding of many-body physics and identify unique properties which may be useful in novel applications such as valleytronics and spintronics.

A trion is formed when either an electron or hole binds to an exciton.  In MoSe$_2$ and WSe$_2$, both positively charged and negatively charged trions have been observed\cite{Ross2013,Jones2013}, however only negatively charged trions have been observed in MoS$_2$.  This difference is attributed to the high unintentional doping of MoS$_2$ that typically leaves samples with an electron density near $10^{13}$ cm$^{-2}$\cite{Mak2013a}; too large to be completely neutralized via electronic back-gating on SiO$_2$/Si$^{++}$ substrates.  These excess electrons greatly increase the likelihood that an exciton and electron meet and bind into a trion, which gives rise to the large trion population in photoexcited MoS$_2$.

PL has been a key characterization tool for studying few layer MoS$_2$ by providing a fast, non-destructive technique for determining the number of layers in a flake\cite{Mak2010c}.  Valley selectivity of optically generated excitons and trions has been demonstrated using polarization-resolved PL\cite{Mak2012}, and momentum-resolved PL has shown that bright excitons have dipole moments solely within the MoS$_2$ plane\cite{Schuller2013}.  Several experiments have monitored trion population via PL while electrically\cite{Mak2013a} or chemically doping MoS$_2$\cite{Mouri2013b}. Recent ultra-fast THz transmission measurements of MoS$_2$ found that the larger mass of the trion relative to the electron results in negative photo-conductivity\cite{Lui2014}.  Importantly, the negative photo-conductivity allowed \citet{Lui2014} to separate the trion contribution from the electron contribution to photo-conductivity, providing direct evidence of trion transport.  The combination of trions being charged as well as abundant when MoS$_2$ is photoexcited makes them interesting candidates for both scientific and technological pursuits because it is possible to control trion transport and density.  

In this article we probe the properties of trions by measuring the long low- energy tail of the trion PL spectra over a wide range of temperatures.  The energy tail length depends on two factors: the trion thermal momentum distribution, and the temperature independent bound state wave function.  We vary temperature to change the momentum distribution and thus independently explore these two factors.  

To the best of our knowledge, the temperature dependence of the trion tail length has not been analyzed in detail for any material.  The closest work is that done by \citet{Ross2013}.  They observed the trion tail in MoSe$_2$ for temperatures below 70 K, which is where the tail is shortest and the peak almost symmetric, but analyze the temperature dependence of the trion tail length.  We also show for the first time that the energy with highest trion PL intensity does not correspond with the zero momentum trion energy, $E^0_{tr}$, and that the PL peak offset from $E^0_{tr}$ changes with temperature.  The peak offset from $E^0_{tr}$ results from the increasing asymmetry of the trion peak as more and more non-zero momentum trions contribute to the PL.  Our analysis allows us to correct for such peak shifts, and if doping were known and defects minimal, would allow us to directly extract the trion binding energy from the PL spectrum.  Thus this work not only provides insight into trions in MoS$_2$, but also establishes a new technique and analysis for understanding trions in general and clarifies important concepts underlying trion photoluminescence.

To study the temperature dependence of the tail length, we prepared monolayer MoS$_2$ via standard mechanical exfoliation from bulk crystals obtained from SPI Supplies.  The sample was prepared on a substrate of degenerately doped silicon wafer with 300 nm of thermal oxide for ideal optical contrast\cite{Benameur2011a,Castellanos-Gomez2010a}.  After exfoliation, monolayer samples were identified optically and verified via PL and Raman.  Spectra were obtained at temperatures ranging from 83 K to 473 K in steps of 25 K using a Linkham THMS 600 cryostat, and we excited our samples using the 514.5 nm line of an Argon ion laser at 250 $\mu$W in a diffraction limited spot. Further details of our sample preparation and measurement methods can be found in the Supporting Information.

\begin{figure}
\includegraphics[width=8.46cm]{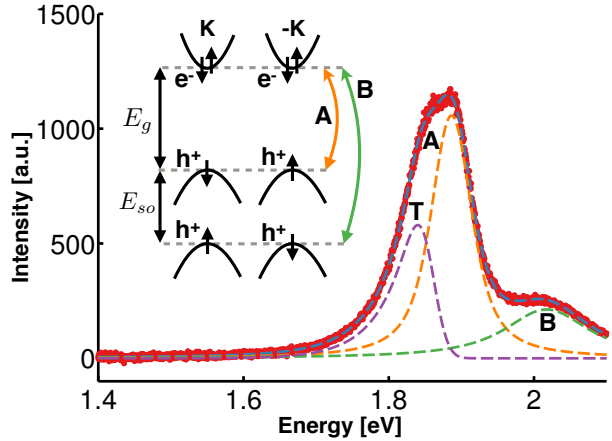}
\caption{\label{band}PL spectrum of MoS$_2$ at 273 K (background subtracted).  Raw spectra are included in Supporting Information.  A and B denote excitons, and T denotes the trion.  Note the trion asymmetric shape with a characteristic low-energy tail.  \textbf{Inset:} Band structure near the K and -K points of the Brillouin zone with band gap, $E_g$, $\sim$1.9 eV, and spin-orbit splitting, $E_{so}$, $\sim$160 meV.}
\end{figure}

Figure \ref{band} shows a typical PL spectrum of MoS$_2$ measured at 273 K together with the fitted components for the excitons and the trion.  As indicated in the inset, the valence band is split by spin-orbit coupling leading to a lower energy A exciton and higher energy B exciton.  While the A and B excitons have Lorentzian peaks, the trion has an asymmetric peak with a long, low-energy tail.

\begin{figure}
\includegraphics[width=17cm]{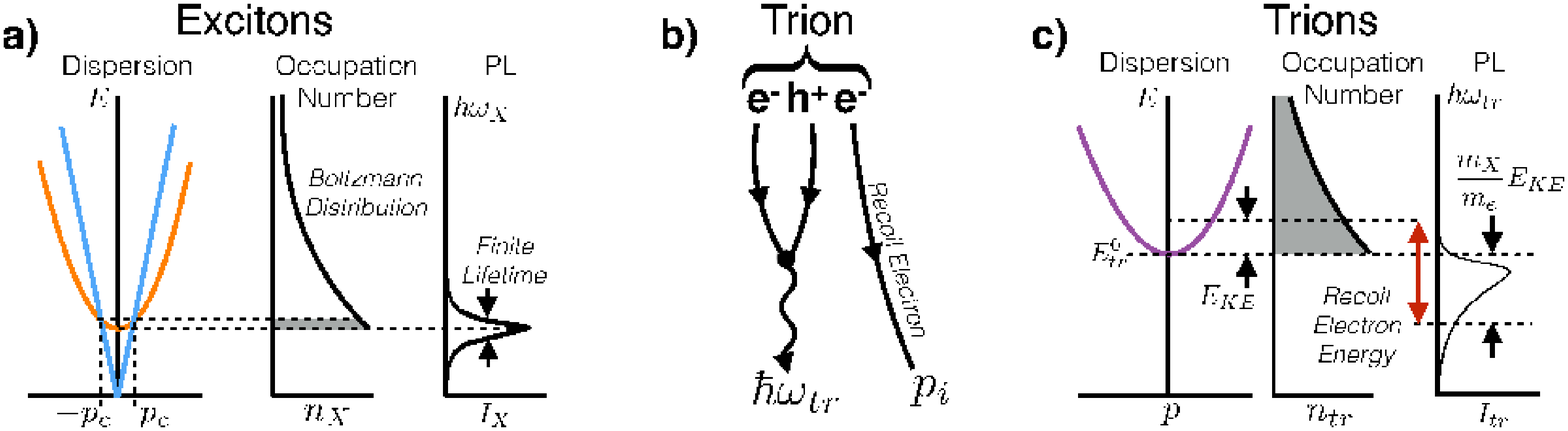}
\caption{\label{DecayPop}  \textbf{a)} The exciton dispersion is shown in orange and photon dispersion in blue with exaggerated momentum, $\sim$200$\times$, to make the light cone visible.  Only excitons within the light cone, $\lvert p \rvert < p_c$, can radiatively decay.  This population is highlighted in the narrow, delta function like region in the occupation number plot, which results in Lorentzian shaped PL.  \textbf{b)} The Feynman diagram for trion radiative decay.  One of the electrons recombines with the hole to emit a photon while the other electron is kicked out to conserve energy and momentum.  \textbf{c)} The trion dispersion is shown in purple with the zero-momentum and kinetic energies of a trion denoted. All trion states can radiatively decay, so that all states in the occupation plot are allowed optical transitions.  To convert from the occupation distribution to the PL, we must account for the energy of the recoil electron resulting in the long low-energy tail.}
\end{figure}

The key to separating the trion and exciton contributions to the PL is accounting for the non-Lorentzian peak shape of the trion, whose asymmetry is due to radiative decay of trions with appreciable momentum, distinctly different from exciton decay.  When an exciton decays, all of its momentum must be carried away by the emitted photon, so only excitons within the light cone, $\lvert p \rvert < p_c$, can radiatively decay, see Figure \ref{DecayPop}a.   This small population of excitons appear like a delta function in occupation space, which when convolved with a Lorentzian to account for the phenomenological finite lifetime creates the exciton Lorentzian PL peak.  Trions, on the other hand, eject an electron when they radiatively decay as shown in Figure \ref{DecayPop}b.  The recoil electron carries away all of the trion's momentum, allowing all trions to decay radiatively.  The corresponding PL spectrum resulting from a Boltzmann distribution\footnote[2]{Our use of the Boltzmann distribution is well supported by thermalization time and lifetime measurements.  Ultra-fast pump-probe experiments find the carrier thermalization time in MoS$_2$ to be $\sim$20 fs\cite{Nie2014a}, while THz pump-probe spectroscopy shows the combined trion non-radiative and radiative lifetime to be $\sim$30 ps\cite{Lui2014}. Since these two time scales are three orders of magnitude different, our assumption of thermal equilibrium is well founded. Further, estimates of trion density based on laser power, spot size and quasi-particle lifetimes show that in our experiment we will not achieve trion densities sufficient for quantum degeneracies, supporting our use of the Maxwell-Boltzmann distribution for trion momentum.} of trion kinetic energies is shown in Figure \ref{DecayPop}c.  Including the energy of the recoil electron is essential when determining the PL spectrum line shape from the distribution of radiatively decaying trion states.  The emitted photon energy is given by $\hbar\omega_{tr} = E^0_{tr} - \frac{m_X}{m_e}E_{KE}$, where $E^0_{tr}$ is the zero-momentum trion energy, $m_X$ is the exciton mass, $m_e$ is the effective electron mass, and $E_{KE}$ is the trion kinetic energy.  Thus the PL spectrum is the thermal trion population distribution ``flipped" over the zero-momentum trion energy and magnified by the ratio $m_X/m_e$ as shown in Figure \ref{DecayPop}c.  The flipping means that the higher energy, non-zero momentum trion states create a \emph{low} energy tail containing information about the thermal distribution and effective hole to electron mass ratio.

To accurately analyze the trion PL shape we need to account for the fact that trions with different momenta will decay at different rates.  This effect is accounted for in the optical matrix element, $M\left(p\right)$, which is a function of the trion momentum, $p$.  Based on theory as well as experimental observations in GaAs quantum wells and TMDCs, it is known that the optical matrix element is well approximated by an exponential of the trion kinetic energy\cite{Esser2000,Ross2013,Stebe1998}.  In the limit that the trion wave function takes the form of a Gaussian wave packet, this approximation becomes exact (see Supporting Information).  With that insight in mind, we approximate the optical matrix element, $M\left(p\right)$, as $M\left(p\right)\propto\exp\left[-\left(\frac{m_X}{m_{tr}}\frac{a}{\hbar}\right)^2p^2\right]$ where $a$ is the standard deviation of the Gaussian wave packet, which we interpret as the effective size of a trion.  Details of the optical matrix element calculation are included in the Supplementary Information.

Accounting for the optical matrix element and the thermal distribution of momenta, the trion PL intensity is given by
\begin{subequations}
\begin{align}\label{PTp}
I_{tr}\left( \hbar\omega \right) & = \exp\left[-\left(E^0_{tr} - \hbar \omega\right)/\epsilon\right] \Theta\left(E^0_{tr} - \hbar \omega\right)/\epsilon\\
\frac{1}{\epsilon} & = \frac{m_e}{m_X}\frac{1}{k_B T} + \left(\frac{m_X}{m_{tr}}\right)^2 \frac{4 m_e a^2}{\hbar^2} \label{Eps}
\end{align}
\end{subequations}
where $\Theta$ is the unit step function, $T$ is temperature, $k_B$ is Boltzmann's constant, and $\epsilon$ is the length, in units of energy, of the low-energy tail of the trion PL.  The spectrum described by eq 1a is normalized to have a total integrated area of 1. It does not include phenomenological broadening, which we include by convolving eq 1a with a Lorentzian of width $\Gamma$.  The first term in eq 1b for $1/\epsilon$ comes from the Boltzmann distribution of trion momenta, and is temperature dependent.  The second term comes from the optical matrix element and is temperature independent.  At low temperatures, the Boltzmann term will dominate and the tail length will be small with a nearly symmetric peak shape.  As temperature is increased, the temperature independent term will dominate, and the energy tail length will increase and finally saturate at a value dictated by the size of the trion.  By measuring the tail length at different temperatures we can separate these two terms.  From the temperature dependent term we can calculate the ratio of the effective hole to electron mass, and from the temperature independent term we can determine the effective size of the trion multiplied by the effective electron mass.

\begin{figure}
\includegraphics[width=7cm]{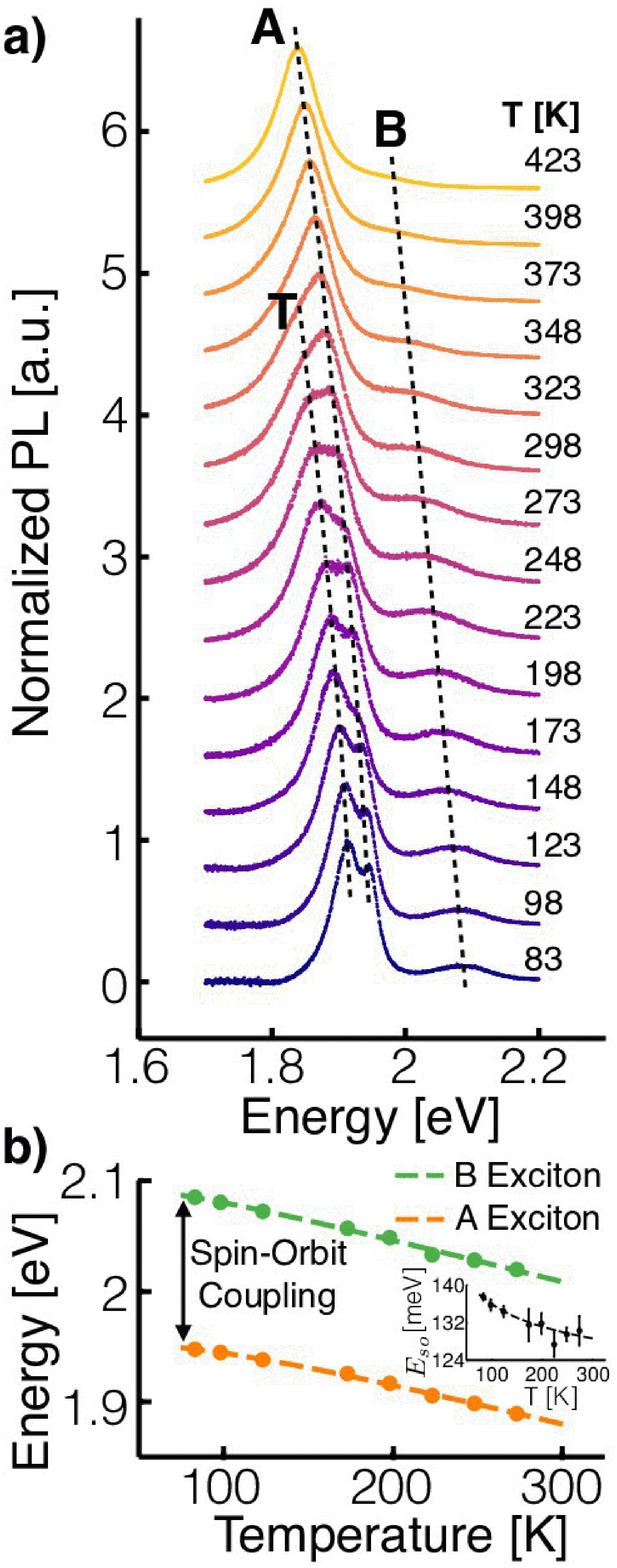}
\caption{\label{PLvsTemp}  \textbf{a)} Monolayer MoS$_2$ PL spectra measured from 83 K to 423 K.  The background has been removed from the spectra, and the spectra have been normalized to make features easily visible.  Guides to the eye for the trion, A exciton and B exciton peak positions are included for clarity.  \textbf{b)} A exciton and B exciton peak positions versus temperature and best-fit to a semi-empirical model\cite{Tongay2012a,ODonnell1991} as described in the text reveals temperature dependent spin-orbit coupling. For most data points the error bars are smaller than the symbols. \textbf{Inset:} Temperature dependent spin-orbit splitting between A and B exciton energies.}
\end{figure}

The measured spectra shown in Figure \ref{PLvsTemp}a are in good agreement with our qualitative expectations.  At higher temperatures thermal excitations reduce the populations of bound stats and shorten the trion and exciton lifetimes resulting in a less prominent, broader trion peak until it completely disappears at $\sim$348 K.  Additionally, we see that  the exciton and trion peaks red-shift as temperature increases because of thermal expansion of the lattice constant\cite{Tongay2012a,ODonnell1991}.  By fitting the spectra (details below) we find the temperature dependence of the A and B excitons to be well described by a semi-empirical model based on electron-phonon coupling\cite{ODonnell1991}, see Figure \ref{PLvsTemp}b.  In this model, the exciton energy is
\begin{equation}
\label{EgVsT}
E_X\left(T\right) = E_X^0 - S \langle \hbar \omega \rangle \left[ \coth \left( \langle \hbar \omega \rangle / 2 k_B T \right) - 1 \right]
\end{equation}
where $E_X^0$ denotes the zero temperature exciton energy, $\langle \hbar \omega \rangle$ represents the average phonon energy contributing to the temperature change of the exciton energy, and $S$ is the effective electron-phonon coupling constant.  Best-fit values for $E_X^0$, $\langle \hbar \omega \rangle$, and $S$ for the A and B excitons are shown in Table \ref{EvsT} along with results from measurements done before the trion was discovered\cite{Tongay2012a}.  There is good agreement with the previous measurements with the exception of the zero temperature energy.  Given that the trion was not known at the time of the earlier work, we suspect that the low temperature spectra were fit to the trion peak instead of the A exciton peak, which would account for the significant difference in zero temperature energy.  Recalling that the A and B excitons are split by spin-orbit coupling, and noting that $\langle \hbar \omega\rangle$ and $S$ are different for the A and B exciton, we conclude that the spin-orbit splitting of the valence band is slightly decreasing as temperature increases.  We hypothesize that the electron wave function is distorted by thermal expansion changing its orbital character or distance to Mo atoms (responsible for the spin-orbit coupling) or both, and leave further analysis as future work.

\begin{table}
\begin{tabularx}{8cm}{l c c c}
\hline
& {\footnotesize $E_X^0$ [eV]} & {\footnotesize  $\langle \hbar \omega \rangle$ [meV]} & {\footnotesize $S$ [-]} \\
\hline\hline
{\footnotesize A} & {\footnotesize 1.952 $\pm$ 0.003} & {\footnotesize 23 $\pm$ 6} & {\footnotesize 2.2 $\pm$ 0.3}\\
{\footnotesize B} & {\footnotesize 2.094 $\pm$ 0.005} & {\footnotesize 16 $\pm$ 6} & {\footnotesize 2.3 $\pm$ 0.2}\\
\hline
{\footnotesize A\cite{Tongay2012a}} & {\footnotesize 1.86} & {\footnotesize 22.5} & {\footnotesize 1.82}
\end{tabularx}
\caption{\label{EvsT} Table of fit parameter values describing the temperature dependence of exciton energy using eq 2.  The bottom row shows previous measurements done without accounting for the trion contribution to the PL\cite{Tongay2012a}.}
\end{table} 

We extract quantitative information from our data by fitting the spectra to two Lorentzian peaks for the A and B excitons, and a long-tailed peak for the trion using eq 1a convolved with a Lorentzian to account for the finite lifetime of the trion.  In addition, we account for a defect peak seen primarily at low temperatures\cite{Mak2013a} with a Gaussian.  We also accounted for inhomogeneous broadening\cite{Mak2013a,Nan2014a,Buscema2014} by convolving the spectra with a Gaussian and performed a detailed noise characterization (see Supporting Information) to properly account for measurement uncertainty.  We have focused our efforts on the data below 300 K, because at higher temperatures the trion peak is much weaker and thermal broadening obscures the trion tail.  

It is challenging to extract both position and shape information self-consistently from spectra with multiple overlapping peaks over the entire temperature range.  First, we take advantage of the separation between the low energy defect peak from the other peaks to fit the defect peak and background parameters independently\cite{Bevington2003}.  These parameters are then held constant, while performing an initial fit to the trion and exciton peaks.  For some spectra the peak widths and inhomogeneous broadening extracted at this stage are unphysically small, in which case we replace them with realistic values which are held constant during a refitting of the remaining parameters.  Then a final fit is performed in which all parameters are free to minimize $\chi^2$, yielding the best-fit over all parameters.  This procedure was successful at yielding physical fit parameters for all but one spectrum, measured at 148 K, so we have removed this spectrum from all further analysis.  Finally, we quantify the confidence intervals of our fit parameters by bootstrapping our data to perform a Monte Carlo simulation of our experiment\cite{Press2007}.  Bootstrapping also estimates the distribution of each fit parameter for each spectrum.  In all cases, the distribution contained a single peak clustered around the best-fit value, indicating the robustness of our approach (see Supporting Information).  

\begin{figure}
\includegraphics[width=17cm]{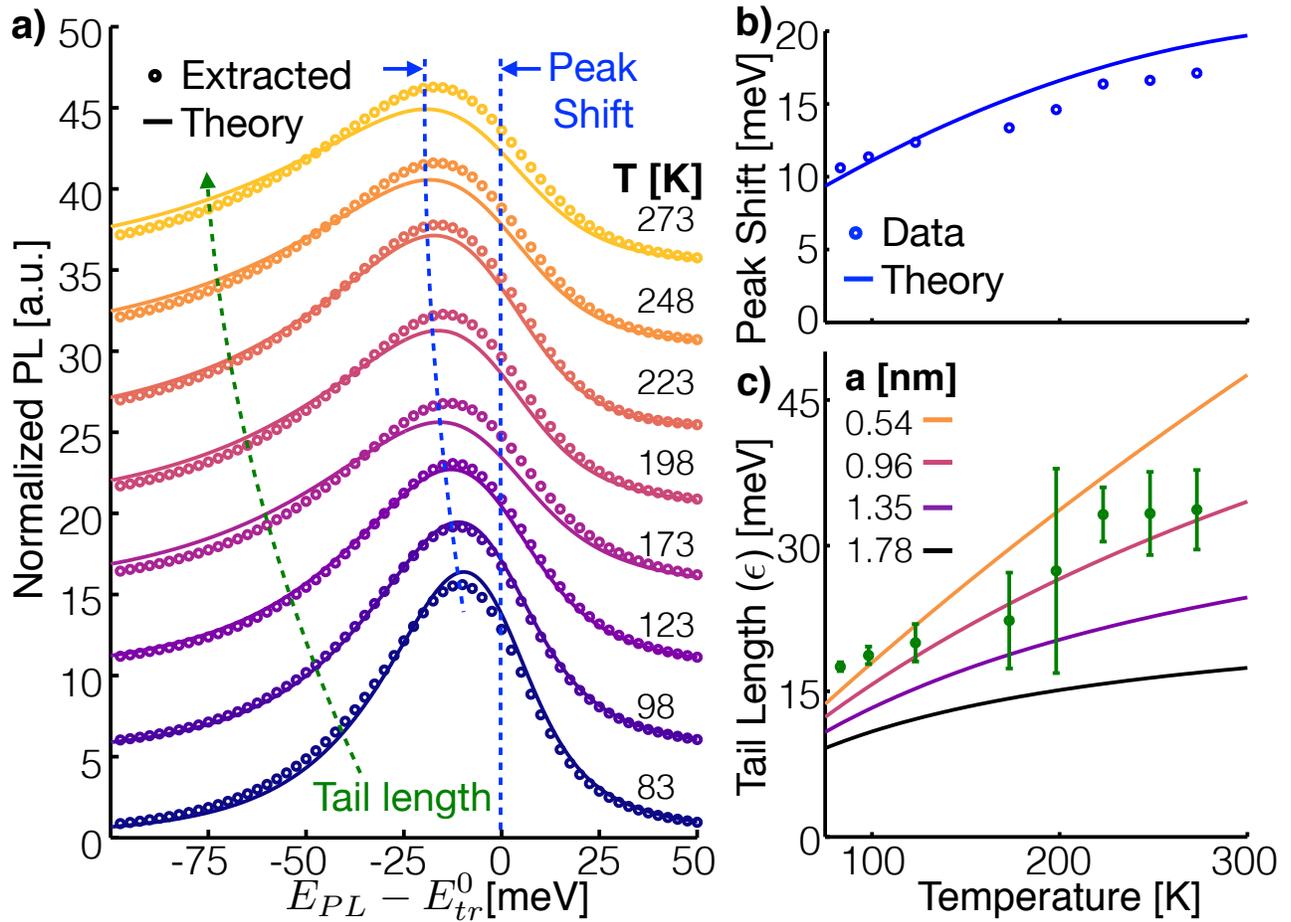}
\caption{\label{LineShape}  \textbf{a)} Trion contribution to the PL as extracted from the data as well as theoretical spectra, using eq 1a, at the same temperatures as our experiments assuming theoretical electron and hole masses\cite{Cheiwchanchamnangij2012}, $a = 0.54$ nm from the best-fit to the data, and phenomenological broadening as measured in our experiments.  As the temperature increases, the peak position red-shifts and the low-energy tail gets longer.  \textbf{b)} Difference between the energy with highest intensity and $E^0_{tr}$ from panel a. \textbf{c)} Extracted trion tail length, $\epsilon$, as well as theoretical curves assuming various values for $a$.}
\end{figure}

The extracted trion contribution to the MoS$_2$ PL spectrum is shown in Figure \ref{LineShape}a along with the expected contribution given the best-fit value for the trion effective size, $a$, and theoretical values for the effective electron and hole masses\cite{Cheiwchanchamnangij2012}.  As temperature increases, the low-energy tail gets longer and the peak red-shifts.  Both of these spectral changes result from the increased population of non-zero momentum trions at higher temperatures.  At low temperatures there are few trions with non-zero momentum, and the spectrum described by eq 1a is nearly a delta function.  This results in a short trion tail, and a highly symmetric PL spectrum peaked near $E^0_{tr}$ when convolved with a Lorentzian to account for the trion finite lifetime.  At higher temperatures, there are more non-zero momentum trions so that the spectrum of eq 1a is broadened asymmetrically on the low energy side.  This gives rise to a long, low-energy tail and a red-shifted peak position after convolution with a Lorentzian to account for broadening.  The red-shift of trion PL peak position with temperature has been observed before without analysis in GaAs quantum wells\cite{Esser2000a}.

It is tempting to take the difference between the trion peak position and exciton peak position as the trion binding energy, however we demonstrate below that this will give an erroneously large trion binding energy. To emphasize this point we have plotted in Figure \ref{LineShape}b the difference between the energy with the highest PL intensity and $E^0_{tr}$ as calculated from the curves in Figure \ref{LineShape}a.  This shows that $E^0_{tr}$ would be measured incorrectly by as much as 20 meV by ignoring the asymmetric trion peak shape. The electron doping dependence of the trion binding energy and Pauli-blocking, which we discuss below, cause additional shifts to the trion peak position\cite{Mak2013a,Zhang2014}

Figure \ref{LineShape}c shows our extracted trion tail lengths, which increase with temperature as expected.  Fitting the tail lengths to eq 1b, the best-fit value for the effective trion size, $a$, is 0.54 nm assuming theoretical values for the effective electron and hole masses\cite{Cheiwchanchamnangij2012}.  We suspect that our two lowest temperature data points are slightly skewed upwards because these spectra had the largest defect peak, and that these data points have driven the best-fit value for $a$ too low.  This possibility is qualitatively supported by the good agreement of our data, excluding the two lowest temperature points, with the 0.96 nm curve.  For comparison we have included in Figure \ref{LineShape}c tail length versus temperature curves for several different values of $a$.  The topmost curve with $a$ = 0.54 nm is the best-fit line to the data.  The curves with $a$ equal to 0.96 and 1.35 nm sizes are derived from absorption spectroscopy experiments on quartz substrates with electron densities of 2$\times10^{12}$ cm$^{-2}$ and 4$\times10^{12}$ cm$^{-2}$ respectively\cite{Zhang2014}.  The curve with size $a$ = 1.78 nm is derived from the theoretical calculations for MoS$_2$ in vacuum with zero electron doping\cite{Berkelbach2013}.  The derived values of $a$ were determined by minimizing the difference between the optical matrix elements of the Gaussian wave packet approximation we use and the more complicated wave packets used in the absorption\cite{Zhang2014} and theory\cite{Berkelbach2013} references (see Supporting Information for details).
 
The difference in the effective trion sizes discussed above highlight the important role of substrate dielectric and electron density on the effective trion size.  In our discussion, we will simplify screening effects by exploring what happens to excitons, as the same arguments apply to trions.  Considering an exciton in three dimensions (3D) as exemplified by a hydrogen atom, all constants in the hydrogen atom Schr\"odinger equation can be removed by nondimensionalization (\emph{i.e.}, when an atom is placed in a dielectric environment, the new electron wave function is simply a rescaled version of the vacuum solution).  However, in a 2D system with dielectric environments above and below that differ from the in-plane dielectric, the proper interaction potential introduces a new length scale\cite{Keldysh1979,Cudazzo2011} which makes it impossible to remove all constants from the 2D exciton Schr\"odinger equation via nondimensionalization.  As a result, changing the substrate dielectric results in a new electron wave function that is not simply a rescaled version of that found in vacuum.  Hence, it is difficult to compare effective trion sizes for samples with different substrate dielectrics.  Since quartz and SiO$_2$ have nearly identical dielectric constants, our results are comparable with the absorption measurements\cite{Zhang2014}, but not with the theoretical calculations\cite{Berkelbach2013} that assume vacuum and zero electron density, and thus a much larger theoretical effective trion size.

Changing the electron density will primarily change the trion size by modifying the MoS$_2$ dielectric function, which in principle leads to the same challenges as accounting for changes to the substrate dielectric.  However, \citet{Zhang2014} used the proper interaction potential in their analysis and found that as doping increases, the exciton and trion radii decrease.  This suggests that the relatively small effective trion size favored by our data could also be the result of a higher doping level than in \citet{Zhang2014}'s samples.  

We note that there are some contradictions in the literature regarding the effect of doping on the dielectric function in monolayer TMDCs.  A recent experiment on monolayer MoS$_2$ treated with gold nanoparticles showed that the exciton PL and absorption peaks red-shift\cite{Exciton2015} due to charge transfer from the nanoparticles to MoS$_2$.  They use the Drude model to show that the doping \emph{decreases} the dielectric function, causing the exciton binding energy to increase, which results in a red-shift to the exciton peak.  In short, they found that increasing doping decreases the dielectric function and increases the exciton binding energy, which is in qualitative agreement with \citet{Zhang2014}'s analysis showing that exciton and trion radii decrease with increasing doping.  This result is in contrast with recent observations of excitons in back-gated monolayer WS$_2$\cite{Chernikov2015} where increasing the electron density was shown to blue-shift the exciton peak.  \citet{Chernikov2015} attribute the blue-shift to Pauli-blocking and increased dielectric screening.  It is important to note that these are two different material systems, MoS$_2$ and WS$_2$, and that band gap renormalization could be dramatically different in these systems.  Given the high unintentional doping of MoS$_2$ on SiO$_2$ substrates our results are in agreement with the MoS$_2$ studies\cite{Zhang2014,Exciton2015}. 

\begin{figure}
\includegraphics[width=8cm]{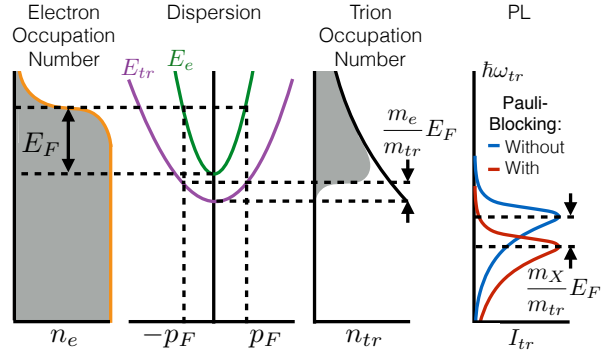}
\caption{\label{PB} The electron occupation number is determined by the Fermi-Dirac distribution with Fermi energy, $E_F$, measured relative to the bottom of the conduction band.  In the dispersion graph the conduction band is drawn in green and the trion dispersion in purple.  Trions with momentum less than the Fermi momentum, $p_F$, cannot decay, which limits the trion radiative states to those with energy above $\frac{m_e}{m_{tr}}E_F$ and red-shifts the trion PL spectrum by $\frac{m_X}{m_{tr}}E_F$.}
\end{figure}

Missing from our discussion of trion decay thus far is Pauli-blocking\cite{Mak2013a,Zhang2014}.  We expect Pauli-blocking to play a role in our experiments because the high unintentional doping of MoS$_2$ exfoliated on SiO$_2$ places the Fermi surface at, or near, the bottom of the conduction band (see Supporting Information). The effect of Pauli-blocking on the trion PL spectra is to red-shift the peak and distort its shape as shown in Figure \ref{PB}.  When a trion decays, the recoil electron must join the Fermi sea.  In the zero temperature limit, this is only possible when the trion's momentum is greater than the Fermi momentum, which is equivalent to limiting the radiative trion states to those with energy above $\frac{m_e}{m_{tr}}E_F$.  Because of the recoil electron's energy this will red-shift the trion PL spectrum by $\frac{m_X}{m_{tr}}E_F$.  We note that red-shifting of the trion peak with electrostatic back-gating has been observed in WSe$_2$\cite{Jones2013}, but Pauli-blocking can only account for about half of the observation (see Supporting Information).  The remainder of the red-shift is likely due to a combination of  the quantum-confined Stark effect\cite{Miller1984} as proposed by \citet{Jones2013}, and many-body effects.

In the zero-temperature limit, the trion PL is only red-shifted by Pauli-blocking, but the shape is undistorted.  However, as temperature increases the sharp cutoff at $\frac{m_e}{m_{tr}}E_F$ in the trion occupation distribution becomes smoothed over an energy range of $\sim\frac{m_e}{m_{tr}} k_B T$, and will distort the high energy side of the trion PL.  When $\frac{m_X}{m_{tr}}k_B T \approx \epsilon$, both the low and high energy sides of the trion PL will be of similar width, and the trion PL spectrum will become symmetric.  If our sample is as heavily doped as seen in the literature, then at high temperatures, $\sim$300 K, it would appear as if the trion PL was symmetric without a low-energy tail because of Pauli-blocking.  We are actively exploring the effects of electron density on trion PL by experimenting with back-gated samples.

In this article we have discussed the physical mechanism through which non-zero momentum trions can radiatively decay and shown how this analysis accurately predicts the resulting asymmetric PL spectrum.  By accounted for the asymmetric trion spectrum, we separated the trion PL from the exciton PL with high precision over a wide range of temperatures, enabling us to estimate the effective size of a trion, and measure the temperature dependence of the A and B exciton energies accurately, resolving the temperature dependent spin-orbit coupling for the first time.  We find that our trion size is consistent with doped MoS$_2$ as measured using absorption spectroscopy.  We have further shown that the zero momentum trion energy, $E^0_{tr}$, will be erroneously determined when using a symmetric, Lorentzian peak, which will result in over-estimating the trion binding energy.  Our model can be used to analyze trions in other systems such as MoSe$_2$ and WSe$_2$ and applied to heterostructures of TMDCs where only the interlayer excitons\cite{Rivera2015} have been investigated.   For interlayer trions, measuring the hole to electron mass ratio via the tail length as presented here would provide significant insight into which material donates the hole and which donates the electrons.  Accounting for the trion tail to accurately separate the trion PL from the exciton PL may also find application in probing the trion contribution to valley and spin hall effects in TMDCs.  Lastly, the signatures of Pauli-blocking discussed in this paper suggest that trions can be stabilized by heavily doping TMDCs.  Given the high unintentional doping of MoS$_2$, Pauli-blocking should be easily achieved in back-gated samples, which leads to several interesting extensions of this research, such as generating a degenerate trion gas or probing the internal orbital degree of freedom such as was recently done with excitons in WSe$_2$\cite{Poellmann2015}.  

\acknowledgement
Author JWC thanks the Department of Defense (DoD), Air Force Office of Scientific Research for its support through the National Defense Science and Engineering Graduate (NDSEG) Fellowship, 32 CFR 168a. This work was also supported by the National Science Foundation Division of Materials Research under grant number 1411008.

\begin{suppinfo}
Raw PL spectra, optical matrix element calculations and adaptations, sample preparation and characterization, CCD noise characterization, Monte Carlo distributions, 2D degenerate gas critical density calculation, and back-gating dependence of trion red-shift due to Pauli-blocking. This material is available free of charge via the Internet at http://pubs.acs.org.
\end{suppinfo}

\beginsupplement
\section{Supporting Information}
\subsection{Raw PL Spectra}
\begin{figure}
\includegraphics[width=\textwidth]{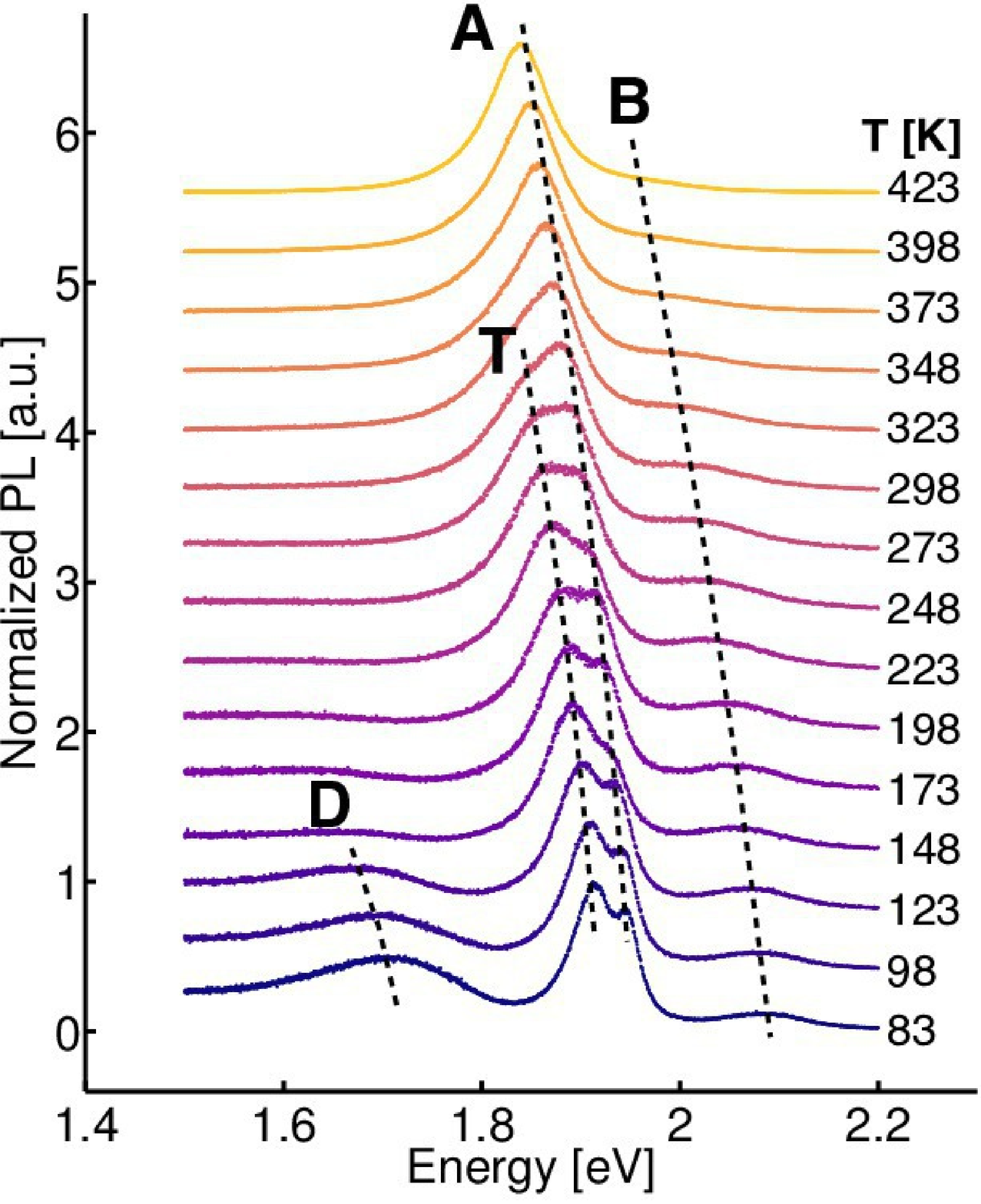}
\caption{\label{RawData} Waterfall plot of data before removing the defect peak, marked with \textbf{D}, and an exponential tail from lower energy defect states.}
\end{figure}

The raw spectra used in our analysis are shown in Figure \ref{RawData} with the defect peak at low temperatures clearly noted.

\subsection{Optical Matrix Element Calculations and Adaptations}
The optical matrix element, $M\left(\mathbf{p}\right)$, which describes the momentum dependent probability of radiative decay of a trion is given by\cite{Esser2000}
\begin{equation}
\label{Mp}
M\left(\mathbf{p}\right) \propto \int \mathrm{d}^2 \mathbf{\rho} \, \psi_{tr}\left(\boldsymbol{\rho}_1=\mathbf{0},\boldsymbol{\rho}_2=\boldsymbol{\rho}\right) \exp\left(-i\frac{\mathbf{p}\cdot\boldsymbol{\rho}}{\hbar} \frac{m_X}{m_{tr}}\right).
\end{equation}
where $\psi_{tr}$ is the trion wave function, and $\boldsymbol{\rho}_1$ and $\boldsymbol{\rho}_2$ are the locations of the electrons relative to the hole.
The matrix element is computed with one of the electrons having a relative coordinate of zero, which makes intuitive sense as one of the electrons should be recombining with the hole at the origin.  Hence, the optical matrix element is just the Fourier transform of the trion wave function's second electron position with the first electron position set to zero.  If we make the substitution into eq S1 that the trion wave function is a Gaussian packet with standard deviation $a$, $\psi^{\mathrm{1P}}_{tr} \left( \mathbf{0},\boldsymbol{\rho} \right) \propto \exp\left(\frac{-\rho^2}{4 a^2}\right)$, then $M^{\mathrm{1P}}\left(\mathbf{p}\right) \propto \exp\left[ - \left( \frac{m_X}{m_{tr}} \frac{a}{\hbar}\right)^2 \mathbf{p}^2\right]$ and we can interpret $a$ as the effective trion size as discussed in the text.  

We have emphasized in the previous paragraph that the Gaussian wave packet corresponds with a one parameter trion wave function by using the superscript $\mathrm{1P}$, the single parameter being the effective trion size, $a$.  However, the absorption spectroscopy work\cite{Zhang2014} and theoretical calculations\cite{Berkelbach2013} both used more realistic two parameter trion wave functions as follows. It is expected, based on observations in GaAs quantum wells, that the two electrons will form a singlet state, in which case $\psi_{tr}$ must be symmetric when swapping the positions of the two electrons.  A standard way to incorporate this symmetry is to form the symmetric product of two single particle wave functions.  In the absorption spectroscopy work\cite{Zhang2014} and theoretical calculations\cite{Berkelbach2013} the single particle wave function is chosen to be the zero angular momentum 2D hydrogen atom wave function
\begin{equation}
\psi_X\left(\boldsymbol{\rho};a\right) = \sqrt{\frac{2}{\pi a^2}}e^{-\rho/a}
\end{equation}
where the $X$ denotes exciton since this is expected to be the lowest energy exciton wave function, and $a$ is a free parameter for the size of the orbital.  The trion singlet state wave function formed from this single particle wave function is
\begin{equation}
\psi_T^{\mathrm{2P}}\left(\boldsymbol{\rho}_1,\boldsymbol{\rho}_2;b,c\right) \propto \psi_X\left(\boldsymbol{\rho}_1;b\right) \psi_X\left(\boldsymbol{\rho}_2;c\right) + \psi_X\left(\boldsymbol{\rho}_2;b\right) \psi_X\left(\boldsymbol{\rho}_1;c\right)
\end{equation}
where the superscript $\mathrm{2P}$ denotes that this is a two parameter wave function with parameters $b$ and $c$ that describe the size of the two electron orbits about the hole. 

To compare results from \citet{Berkelbach2013} and \citet{Zhang2014} with ours we have adapted their values for $b$ and $c$ by find the value of $a$ that minimizes the sum square error (SSE) between the optical matrix elements of their two parameter wave function and our single parameter wave function.  The SSE is given by
\begin{equation}
\mathrm{SSE} = \int \mathrm{d}^2p \left[M^{\mathrm{1P}}\left(\mathbf{p}\right) - M^{\mathrm{2P}} \left( \mathbf{p}\right) \right]^2
\end{equation} 
where $M^{\mathrm{1P}}$ is the matrix element given by our one parameter, Gaussian wave function, and $M^{\mathrm{2P}}$ is the matrix element given by more complicated two parameter, symmetrized hydrogen atom wave function.  The values of $a$ that minimized the SSE for different values of $b$ and $c$ found in the literature are shown in Table \ref{effTrionRadius}.

\begin{table}
\begin{tabular}{l c c c c c c}
\hline
author & Substrate & $n_e$ [cm$^{-2}$] & b [nm] & c [nm] & $\Rightarrow$ & $a$ [nm] \\
\hline
\citet{Zhang2014} & Quartz & $4\times10^{12}$ & 0.83 & 1.08 & $\Rightarrow$ & 0.96\\
\citet{Zhang2014} & Quartz & $2\times10^{12}$ & 0.93 & 1.77 & $\Rightarrow$ & 1.35\\
\citet{Berkelbach2013} & Vacuum & 0 & 1.03 & 2.52 & $\Rightarrow$ & 1.78\\
\end{tabular}
\caption{\label{effTrionRadius} Table of two parameter, $b$ and $c$, wave function sizes along with adapted single parameter wave function size, $a$.}
\end{table}

\subsection{Sample Preparation and Characterization}
Sample substrates were prepared by dicing them into 1cm by 1cm squares, then cleaned with isopropyl alcohol and acetone sonication baths for 10 minutes each, followed by piranha etch for 20 minutes to remove all traces of contamination.  The sample was mounted to the Linkham cryostat using a thin layer of silver paint (PELCO Colloidal Silver) to ensure good thermal equilibrium with the cryostat stage.  To prevent contamination from condensing on our cold sample we created a small chamber within the Linkham cryostat with the bottom being the silicon substrate, sides made of PDMS and sealed on top with 0.17 mm thick cover glass.  

The PL spectra from our samples was collected using backscatter geometry on a Renishaw spectrometer with a home built microscope with extra long working distance Mitutoyo 100x objective (0.7 NA) for compatibility with our cryostat.  The full width half max (FWHM) laser spot size after passing through the windows of the Linkham cryostat was determined to be $\sim$790nm by measuring the Si Raman peak as the beam was scanned over the edge of a gold target.  The spectrometer utilized a 1800 lines per mm grating dispersing the beam onto a CCD with bin sizes smaller than 0.67 meV. The incident light is prepared in linear polarization to create excitons and trions at both the K and -K valleys, and the detection path contained no polarization selective optics other than the intrinsic efficiency of the grating not included in our analysis as we expect populations at the K and -K valleys to be identical due to thermal equilibration.

\subsection{CCD Noise Characterization}
An essential part of $\chi^2$ minimization is determining the noise present in the measurements.  Here the main component of noise comes from the read and shot noise of the CCD.  We've implemented the CCD Transfer Method\cite{Janesick1997} to characterize the noise of our system.  In this technique, the noise in the digital output of the CCD is measured as the intensity of incident illumination is varied.  There are three components to this noise, and each component varies differently as a function of incident illumination.  \textbf{1) Read Noise:} This results from thermal and quantization noise created during conversion from charge to a digital signal when reading out the CCD. Read noise is independent of the intensity of the illumination, so on a log-log plot has a slope of $0$.  \textbf{2) Shot Noise:} Shot noise stems from the fundamental fact that photons and electrons are quantized, and as a result the number of them collected varies with a Poisson distribution.  The standard deviation of a Poisson distribution grows as the square root of the mean value, so on a log-log plot this noise will have a slope of $1/2$.  \textbf{3) Correlation Noise or Fixed Pattern Noise:} It is expected that the photons that hit the CCD at one place and time are independent of photons that hit the CCD at another place or time.  However, there are correlations between the signal measured at different locations due to slight fabrication imperfections.  Further, there can be bleeding of charge from pixel to pixel creating spatial correlation, and charge can be deposited into deep layers of the CCD that don't discharge during the normal measurement process creating time correlations.  These correlations result in a standard deviation that grows linearly with incident illumination, which on a log-log plot will have a slope of $1$.

During normal CCD noise measurements the interest is in determining the noise characteristics of individual pixels.  However when our CCD is used to measure spectra all the charge accumulated in the pixels in a single column are first aggregated and then read out as a single value for each column.  By aggregating the charge first and making a single read the amount of read noise is significantly reduced, but there is noise introduced from aggregating the charge.  To handle this difference properly we have used our hardware to ``capture'' spectra under various illumination intensities, and determined noise characteristics of the columns.

\begin{figure}
\includegraphics[width=\textwidth]{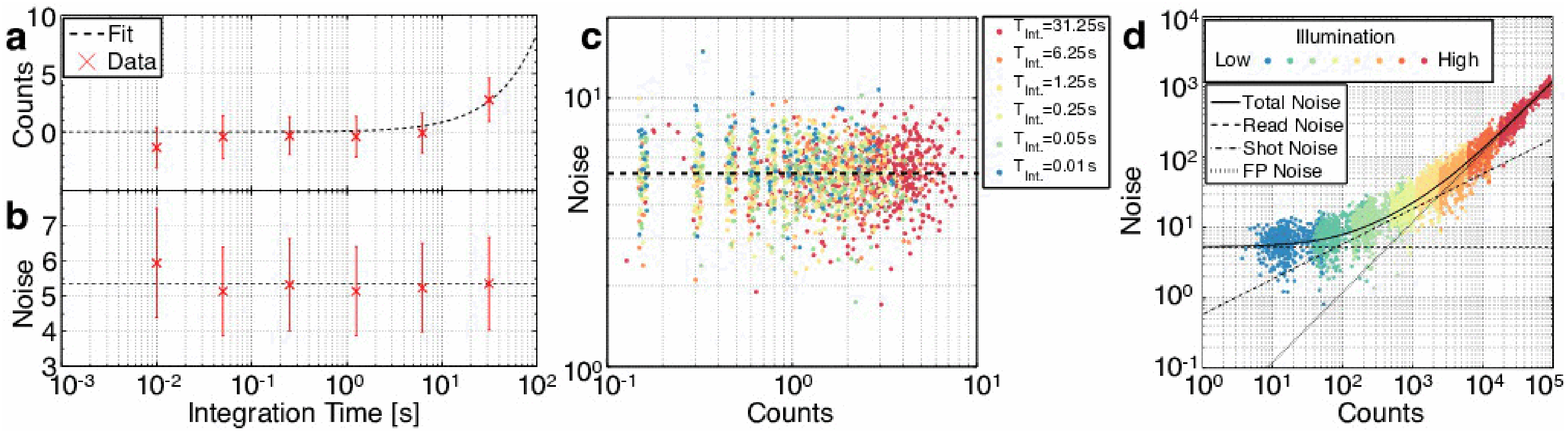}
\caption{\label{CCDNoise}  Measurements of CCD Noise \textbf{a} Column counts measured with various integration times without any illumination on the panel.  There is a small amount of leakage current in the panel on the order of $0.084$ counts per second which results in the small increase in counts at longer integration times.  Error bars represent the standard deviation taken across all columns.  \textbf{b} Column noise versus integration time measured without any illumination on the panel.  The noise is constant showing that the small leakage current does not contribute significant shot noise.  The error bars represent the standard deviation taken across all columns.  \textbf{c} Column noise versus column counts from dark measurements on a log-log plot.  This verifies that noise is read noise, because the slope is $0$. Each point in the graph represents a different column of the CCD and integration time is labeled by color.  \textbf{d} Column noise versus column counts as illumination intensity is varied.  This log-log plot nicely separates noise sources by their differing slopes.  Each point corresponds with a measurement on a single column and intensity with intensity labeled by color.}
\end{figure}

Our measurements are shown in Figure \ref{CCDNoise}.  In panels \textbf{a}, \textbf{b} and \textbf{c} we show dark (no illumination) measurements plotted against integration time.  As seen in panel \textbf{a}, there is a small amount of leakage current on the order of $0.084$ counts per second.  Panel \textbf{b} shows that the small leakage current isn't enough to generate significant shot noise, and we find that the noise is constant as a function of time and gives us a good estimate of $5.4$ counts for the read noise.  To verify that this noise is indeed read noise we have plotted in panel \textbf{c} the noise versus the number of counts on a log-log plot showing it has a slope of $0$ as expected for read noise.  In panel \textbf{d} we have plotted the noise versus counts under various illumination intensities and performed a least squares fit to
\begin{equation}
\sigma = \sqrt{\sigma_R^2 + a N + (b N)^2}
\end{equation}
where $\sigma_R$ is the read noise as determined by the dark measurements, $N$ is the number of counts, $a$ is a conversion constant which accounts for the difference between the digital values measured and the number of electrons detected\cite{Janesick1997}, and $b$ accounts for time correlation within our measurements.  The fit as well as each of the noise components have been plotted in panel \textbf{d} showing the different slopes for each of the noise sources.  We find that our conversion constant, $a$, is $0.337$ counts, which indicates there are approximately $3$ electrons per digital count.  Lastly our fit has determined $b$ to be $0.012$.  In our case the correlation noise is introduced primarily from our poor illumination source, and we neglect the correlation noise term when we calculated the standard deviation of measured spectra.  Explicitly, the standard deviation we associate with the PL intensity measured in bin $i$ is
\begin{equation}
\sigma_i = \sqrt{\sigma_R^2 + a C_i }
\end{equation}
where $C_i$ is the number of counts in bin $i$.  The calculated $\sigma_i$ are essential to the $\chi^2$ minimization process because they properly weight the measurements by their uncertainty.  In our fitting procedure we minimize
\begin{equation}
\chi^2 = \sum_i \frac{\left( C_i - f(\vec{b},E_i)\right)^2}{\sigma_i^2}
\end{equation}
where $\vec{b}$ are the fit parameters, $f$ is the fit function, and $E_i$ is the energy of bin $i$\cite{Bevington2003}. 

\subsection{Monte Carlo Distributions}
Fit errors were determined by bootstrapping the data to create 100 new data sets for each measured spectra.  Each of the 100 new data sets was then fit to create a distribution of fit parameters with different $\chi^2$ values.  Confidence intervals were calculated by ranking fit results by their $\chi^2$ values and increasing the cutoff $\chi^2$ value until 68\% of the fits or more fell between the upper and lower bounds set by the cutoff.  Note that this was necessary as the Jacobian fit matrix was very flat at the bottom of the $\chi^2$ potential, so simple first order propagation of error yielded erroneously large fit errors.

\begin{figure}
\includegraphics[width=\textwidth]{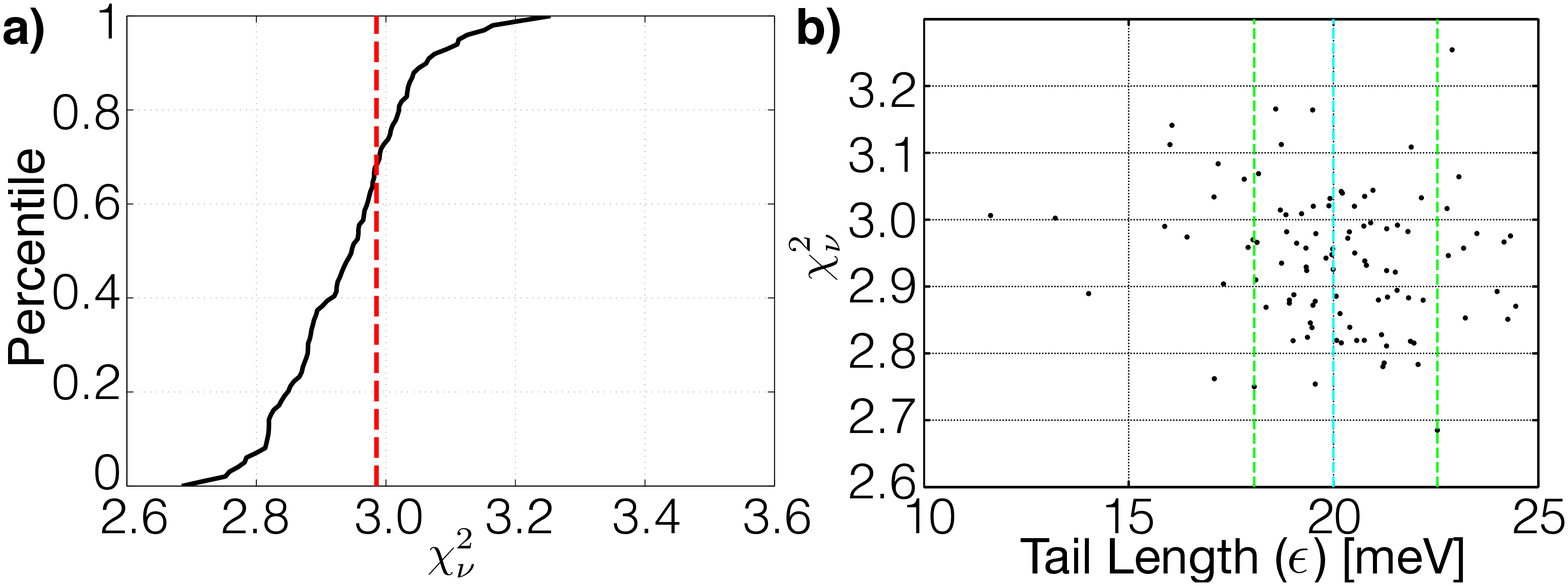}
\caption{\label{MC} \textbf{a)} $\chi^2_{\nu}$ cumulative distribution for PL measured at 123 K. The dashed red line denotes the 68th percentile, which indicates the threshold $\chi^2_{\nu}$ value that sets the 1$\sigma$ confidence interval for the joint probability distribution of all variables.  \textbf{b)} Distribution of trion tail length, $\epsilon$, for the PL measured at 123 K.  The cyan vertical line denotes the best-fit value to the data, and the green vertical lines denote the upper and lower cut offs which contain 68\% of the distribution, establishing the 1$\sigma$ confidence interval for the single variable probability distribution for $\epsilon$.}
\end{figure}

Figure \ref{MC}a shows the cumulative distribution of reduced chi squared, $\chi^2_{\nu}$, for 100 fits to the bootstrapped data from the PL measurements made at 123 K.  The distribution appears like an error function suggesting that 100 samples is a good representative sample.  Figure \ref{MC}b shows the distribution of trion tail lengths.  The best-fit to the original data is shown in cyan, which the distribution is centered on.  That there is no other value of the tail length around which the distribution is clustered is indicative of a robust fit.  The green bars indicate the low and high cutoffs that surround 68\% of the distribution and set the 1$\sigma$ confidence interval for the trion tail length for the data measured at 123 K.  All parameters for each of the PL measurements similarly show distributions clustered around a single value of the fit parameter, indicating a robust fit.

\subsection{2D Degenerate Gas Critical Density Calculation}
\begin{figure}
\includegraphics[width=\textwidth]{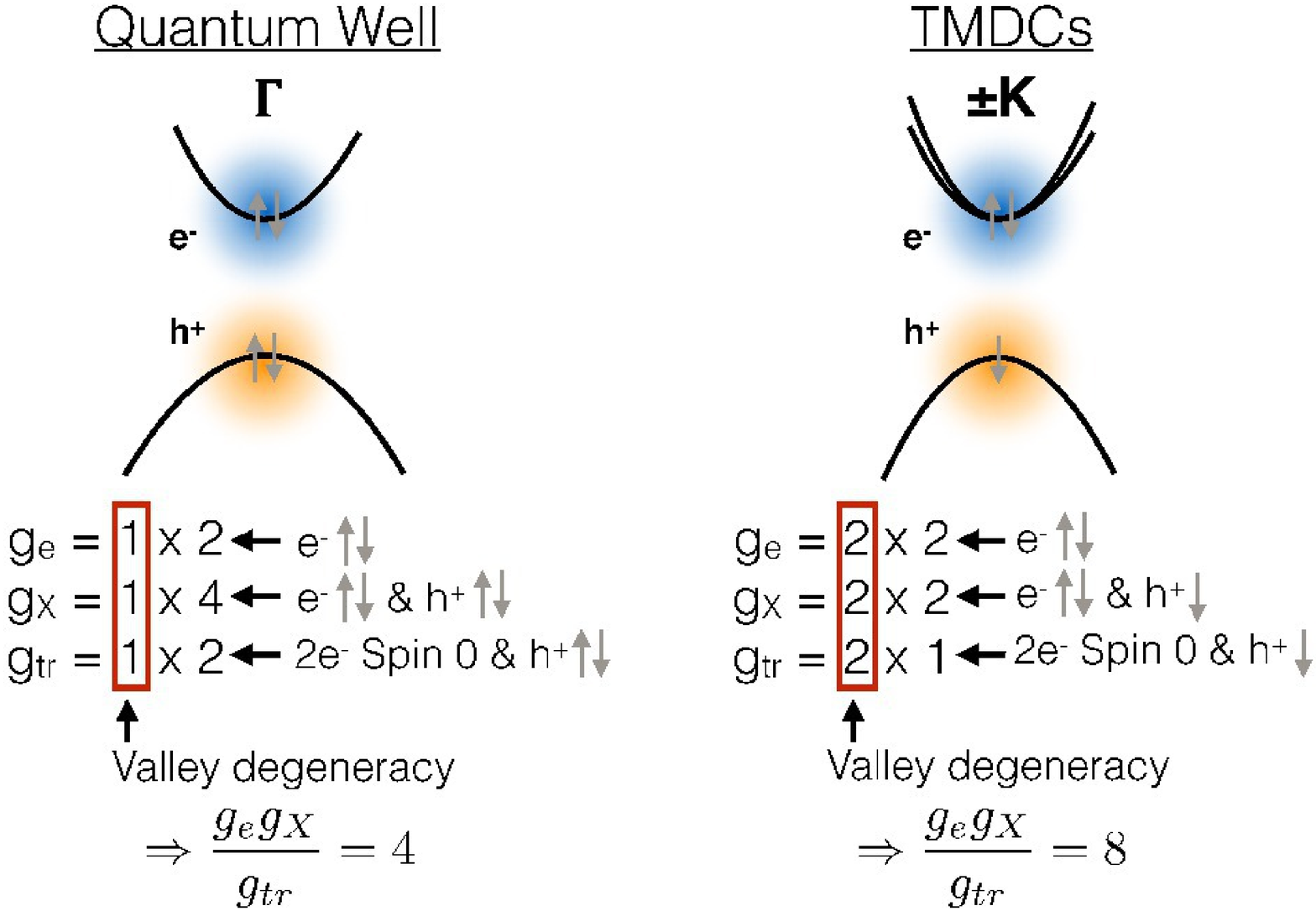}
\caption{\label{DegComp}  Comparison between the degeneracy of electrons, excitons, and trions in GaAs quantum wells and trions in TMDCs.}
\end{figure}

Essential to calculating the critical densities properly is getting the degeneracies correct.  The literature is not clear about the degeneracies of electrons, excitons and trions in GaAs quantum wells in comparison with TMDCs, so we have drawn a diagram and shown the explicit counting in Figure \ref{DegComp} in order to shed light on this topic.  We have also included in Figure \ref{DegComp} the ratio $g_e g_X/g_{tr}$ as this ratio has frequently appeared in the literature when using the mass action law to determine the trion binding energy in GaAs quantum wells \cite{Ron1996,Siviniant1999,Esser2000a,Vercik2002} and MoSe$_2$\cite{Ross2013}.  

The critical density at which quantum statistics become relevant is $n_c = g/\lambda^2$ where $g$ is the degeneracy of the states and $\lambda$ is the thermal de Broglie wave length, $h/\sqrt{2\pi m k_B T}$, where $h$ is Planck's constant, $m$ is the mass of the particle, $k_B$ is Boltzmann's constant and $T$ is temperature.  Table \ref{nCrit} shows the critical densities for electrons, excitons, and trions over the temperature range of our experiments, 80 K to 500 K.  For these calculations we used effective electron and hole masses as calculated via DFT\cite{Cheiwchanchamnangij2012}.

\begin{table}
\begin{tabular}{l c c}
Particle & 80 K & 500 K \\
\hline
Electron & 0.2$\times10^{13}$ cm$^{-2}$ & 1.2$\times10^{13}$ cm$^{-2}$\\
Exciton & 0.1$\times10^{13}$ cm$^{-2}$ & 0.5$\times10^{13}$ cm$^{-2}$\\
Trion & 0.1$\times10^{13}$ cm$^{-2}$ & 0.7$\times10^{13}$ cm$^{-2}$\\
\end{tabular}
\caption{\label{nCrit} Critical densities for the onset of quantum statistics.  All values are in units of $\times10^{13}$ cm$^{-2}$}
\end{table} 

The high unintentional doping of MoS$_2$ on SiO$_2$ is near $10^{13}$ cm$^{-2}$, which is larger than the electron critical density in Table \ref{nCrit} at 80 K and near the critical density at 500 K.  This indicates that Pauli-blocking is likely, and calculations of the trion binding energy using the mass action law should use the quantum statistics in the calculation.

\subsection{Trion Red-Shift Back-gate Dependence}
The 2D electron density in a uniformly back-gated sample is given by
\begin{equation}
n_e = \frac{\varepsilon_{ox} V_{bg}}{t e}
\end{equation}
where $\varepsilon_{ox}$ is the gate oxide dielectric constant, $V_{bg}$ is the back-gate voltage, $t$, is the gate oxide thickness, and $e$ is the electron charge.  In the zero temperature limit the Fermi energy of a 2DEG is
\begin{equation}
E_F = \frac{2 \pi \hbar^2}{g_e m_e}n_e.
\end{equation}
Combining these two equations and noting that the trion red-shifts by $m_X/m_{tr} \times E_F$ we find the slope with which the trion PL spectrum red-shifts due to back-gate voltage to be
\begin{equation}
\text{slope} = \frac{\varepsilon_{ox}}{t e}\frac{2 \pi \hbar^2}{g_e m_e}\frac{m_X}{m_{tr}} = \left[ \text{WSe}^2\text{ on 300 nm SiO}_2 \right] = 0.11 \frac{\text{meV}}{\text{V}} \label{slope}.
\end{equation}

We find the slope to be 0.11 meV / V for WSe$^2$ on 300 nm of SiO$_2$, the geometry of the samples used by \citet{Jones2013}.  In this calculation we used theoretical values for the electron and hole masses of WSe$^2$\cite{Ramasubramaniam2012} as well as $g_e = 4$.  The slope of the trion binding energy versus back-gate voltage in the inset of Figure 2b of reference \cite{Jones2013} is $\approx$0.2 meV / V, so about half of the slope is accounted for by Pauli-blocking.  We believe the remaining portion of the slope, $\approx$0.09 meV / V is due to the quantum-confined Stark effect\cite{Miller1984} as proposed by \citet{Jones2013}.

\bibliography{LongTailedTrions}

\end{document}